\documentclass[prl,twocolumn,notitlepage,superscriptaddress,longbibliography]{revtex4-1}
\usepackage{amsmath,amssymb,bm}
\usepackage{graphicx}
\usepackage{xcolor}
\usepackage{comment}
\usepackage[pdfpagelabels=false,colorlinks=true,urlcolor=teal,citecolor=teal,linkcolor=teal]{hyperref}
\graphicspath{{figures/}}
\allowdisplaybreaks

\newcommand{\LO}{\mathrm{LO}}
\newcommand{\EW}{\mathrm{EW}}
\newcommand{\QCD}{\mathrm{QCD}}

\begin{document}

\title{Full Electroweak Corrections to Higgs Pair Production in Higgs Effective Field Theory}
\author{Jia-Le Ding}
\affiliation{School of Physics, Shandong University, Jinan, Shandong 250100, China}
\author{Jian Wang}
\email{j.wang@sdu.edu.cn}
\affiliation{School of Physics, Shandong University, Jinan, Shandong 250100, China}
\affiliation{Center for High Energy Physics, Peking University, Beijing 100871, China}
\author{ Huai-Min Yu }
\affiliation{INFN, Sezione di Milano-Bicocca, Piazza della Scienza 3, 20126 Milano, Italy}

\begin{abstract}

We present theoretical predictions for Higgs pair production via gluon fusion, including the full electroweak corrections within the Higgs Effective Field Theory framework. In addition to the Higgs self-coupling modifiers \(\kappa_3\) and \(\kappa_4\), we consider the top-Yukawa coupling modifier, the gauge-boson--Higgs coupling modifiers, and the top-pair--Higgs-pair and gluon--Higgs local couplings.
We have developed two different renormalization prescriptions, off-shell and on-shell, to cancel the ultraviolet divergences in the two-loop amplitudes, and have performed the calculation in both the Feynman and unitary gauges.
We find that the electroweak corrections are significant for \(|\kappa_3|>3\) and markedly change the correlation between \(\kappa_3\) and other Higgs couplings. The Higgs-pair invariant mass distributions are sensitive to \(\kappa_3\) and other Higgs couplings, and the electroweak corrections are sizable in many parameter scenarios and kinematic regions.
Our results, combined with the previously obtained QCD corrections, provide precise predictions for the inclusive and differential Higgs-pair cross sections as functions of the coupling modifiers, and could thus help derive more stringent constraints from experimental data.

\end{abstract}
\maketitle

After the discovery of the Higgs boson (\(h\)) at the Large Hadron Collider (LHC) at CERN~\cite{ATLAS:discovery,CMS:discovery}, measurements of its properties and its interactions with gauge bosons and fermions are in broad agreement with the Standard Model (SM)~\cite{ATLAS:portrait,CMS:portrait}. A crucial question nevertheless remains: whether the Higgs self-interactions follow the SM prediction. Determining these interactions and thereby reconstructing the Higgs potential can help us understand the nature of the electroweak phase transition and vacuum stability~\cite{Sher:1989,Djouadi:1999}.
Addressing this question is one of the primary goals of the LHC physics program. Higgs-pair production provides the principal direct probe of the Higgs trilinear self-coupling \(\lambda_3\) at the LHC, while complementary indirect information is available from single-Higgs observables through radiative corrections~\cite{McCullough:2014,Gorbahn:2016,Degrassi:2016,ATLAS:singleDouble,CMS:singleDouble}.
The small cross sections and the limited number of experimentally accessible events make a precise determination challenging. A combination of the ATLAS and CMS searches yields the constraint $-0.71<\kappa_3\equiv\lambda_3/\lambda_3^{\rm SM}<6.1$ at the 95\% confidence level, where $\lambda_3^{\rm SM}$ is the SM Higgs trilinear self-coupling~\cite{ATLASCMS:2026}. 
The high-luminosity LHC is expected to substantially improve this situation: combined ATLAS and CMS projections indicate sensitivity to the trilinear self-coupling at the level of a few tens of percent~\cite{HLLHC:2025}. This prospect calls for increasingly precise theoretical predictions for Higgs-pair production.

Sustained theoretical effort has substantially improved the description of the dominant gluon-fusion Higgs-pair production channel, $gg\to hh$.
Within the SM, QCD corrections have been calculated through next-to-next-to-next-to-leading order (N$^3$LO) in the large-top-quark-mass limit~\cite{Dawson:1998,deFlorian:2013,Grigo:2014jma,deFlorian:2016diff,Chen:2020a,Chen:2020b,Chen:2026zmi}.
Both the next-to-leading-order (NLO) QCD corrections~\cite{Borowka:2016,Borowka:2016full,Baglio:2019,Baglio:2020,Baglio:2020wgt,Bagnaschi:2023HH,Davies:ggxy} and electroweak (EW) corrections~\cite{Bi:2024,Heinrich:2024,Bhattacharya:2026EW,Muhlleitner:2022Yukawa,Davies:2023EWlargeMt,Davies:2026EWhighEnergy,Zhang:2024Factorizable,Davies:2022HiggsYukawa,Davies:2025YukawaSelf,Bonetti:2025light} are available including finite top-quark mass dependence.
Finite top-quark mass effects have also been incorporated into approximate NNLO QCD predictions~\cite{Grazzini:2018}.
These developments provide a precision baseline for testing the Higgs potential.

However, the present data have not established that the Higgs self-couplings match their SM values.
Experimental interpretations must therefore accommodate general self-coupling values and possible correlated modifications of other interactions relevant to  $gg\to hh$.
The relevant precision predictions have been predominantly based on higher-order QCD corrections. For variations in the Higgs trilinear self-coupling, these predictions include N$^3$LO QCD results in the large-\(m_t\) limit~\cite{Chen:2020b} and NLO predictions incorporating finite-\(m_t\) effects~\cite{Baglio:2020wgt,Borowka:2016full,Baglio:2020,Bagnaschi:2023HH,Davies:ggxy}.
More general anomalous interactions have been studied at full NLO in QCD in the Higgs Effective Field Theory (HEFT)~\cite{Buchalla:2018}, including matching to parton showers~\cite{Heinrich:2020}, and at approximate NNLO in QCD~\cite{deFlorian:2021}. Calculations at NLO in QCD within the Standard Model Effective Field Theory (SMEFT) have further addressed the effects of higher-dimensional operator insertions and of the truncation of the EFT expansion~\cite{Heinrich:2022,Heinrich:2023}.

Extending this program to EW corrections for general Higgs couplings is essential. First, the EW corrections may be significant in some kinematic regions, such as large electroweak Sudakov logarithms at high energies, or in certain parameter scenarios. Second, the EW corrections may introduce additional coupling dependence and alter the correlations among couplings.
To this end, several obstacles must be overcome.
Besides the complexity of calculating two-loop integrals involving multiple scales, there is a more fundamental issue of theoretical consistency. In the SM, consistent relations among the Higgs self-interactions, Yukawa interactions, gauge interactions, and their counterterms play a crucial role in ensuring the cancellation of ultraviolet divergences and the gauge independence of the renormalized amplitude. Independently rescaling selected vertices while otherwise retaining the SM interaction and counterterm structure does not, in general, preserve these cancellations; see, e.g., Ref.~\cite{Heinrich:2024}.
A coherent treatment must embed the modified couplings into a gauge-invariant framework and include additional operators up to the chosen order. Both SMEFT and HEFT provide such frameworks, with different symmetry realizations and power counting~\cite{Buchalla:2014,Brivio:2017vri}. Implementing a consistent variation of the Higgs couplings in a loop induced process is a central obstacle to obtaining EW predictions beyond the SM coupling limit.

So far, the available results for nonstandard Higgs self-interactions have covered only selected contributions rather than the full general-coupling EW corrections. Studies of the Higgs quartic self-interaction in SMEFT~\cite{Borowka:2019,Bizon:2019} and calculations of enhanced Higgs self-interaction effects, e.g., the \(\kappa_3^4\) terms, in HEFT~\cite{Li:2024,Ding:report}, have demonstrated the importance of this program. Nonetheless, these results do not capture all terms sensitive to Higgs self-couplings, nor the dependence on Higgs couplings with other particles or the correlations among them.

In this Letter, we present a comprehensive calculation of the EW corrections to gluon-fusion Higgs-pair production in the HEFT. We include the Higgs self-coupling modifiers \(\kappa_3\) and \(\kappa_4\), together with the top-Yukawa coupling modifier \(\kappa_t\), the gauge-boson--Higgs coupling modifiers \(\kappa_V\) and \(\kappa_{2V}\), the top-pair--Higgs-pair local coupling \(c_{2t}\), and the gluon--Higgs local couplings \(c_{ggh}\) and \(c_{gghh}\), through the effective Higgs Lagrangian 
\begin{align}\label{eq:potential}
 \mathcal{L}_{\rm HEFT} &\supset
       \delta t h-\left(m_h^2+\frac{\delta t}{v}\right)\left(\kappa_3\frac{h^3}{2v}
        +\kappa_4\frac{h^4}{8v^2}\right) 
      \nonumber\\
 &-m_t
           \left(\kappa_t\frac{h}{v}+c_{2t}\frac{h^2}{v^2}\right){\bar Q}_L U e_t t_R+{\rm h.c.} \nonumber\\
 &+\frac{v^2}{4}\left(2\kappa_V\frac{h}{v}+\kappa_{2V}\frac{h^2}{v^2}\right){\rm Tr}[D_{\mu}U^{\dagger}D^{\mu}U]\nonumber\\
 &+\frac{\alpha_s}{8\pi}
 \left(c_{ggh}\frac{h}{v}+c_{gghh}\frac{h^2}{v^2}\right)G^a_{\mu\nu}G^{a\mu\nu} 
\end{align}
where $Q_L=(t_L,b_L)^T, e_t=(1,0)^T$,  $U=\exp(i\tau^a\phi^a/v)$ with the Goldstone fields $\phi^a$ and the Pauli matrices $\tau^a$, and $D_\mu$ is the covariant derivative \cite{Buchalla:2014,Buchalla:2014PowerCounting,Buchalla:2018Renormalization}.
The operators in the first three lines belong to the leading-order (LO) Lagrangian of the HEFT, while those in the last line are at NLO according to the power counting of chiral dimensions~\cite{Buchalla:2014PowerCounting,Buchalla:2018}. However, they contribute to the \(gg\to hh\) amplitudes at the same order as the LO operators associated with \(\kappa_3\), \(\kappa_t\), and \(c_{2t}\).
The parameters \(\kappa_4\), \(\kappa_{V}\), and \(\kappa_{2V}\) contribute to the NLO EW corrections to \(gg\to hh\). We have introduced the tadpole parameter \(\delta t\) so that the SM results are recovered when all \(\kappa_i=1\) and \(c_i=0\) after renormalization in the modified minimal subtraction (\(\overline{\mathrm{MS}}\)) scheme. 
The Higgs-pair cross sections thus computed substantially extend the theoretical description required for experimental analyses with general Higgs couplings.

We begin by considering the two-loop amplitudes involving coupling modifiers and additional local couplings. The Feynman diagrams are generated with \texttt{FeynArts}~\cite{FeynArts} using the Feynman rules derived from \(\mathcal{L}_{\rm HEFT}\). We find that five- and six-point vertices, such as \(\bar t t h h \phi^+\phi^-\), can appear in the two-loop diagrams. Therefore, there are 1176 more diagrams than in the SM calculation.
Algebraic manipulation of the amplitudes, including projection to two form factors and the organization of scalar integrals into 116 families, is performed with \texttt{CalcLoop}~\cite{CalcLoop}.
Integration-by-parts reduction and the construction of differential equations for the master integrals in each family are performed with \texttt{Blade}~\cite{Blade}, which internally uses \texttt{FiniteFlow}~\cite{FiniteFlow}.
Boundary values of the master integrals are evaluated using the auxiliary mass flow method~\cite{AMFmethod,AMFmultiloop,AMFlinear} with \texttt{AMFlow}~\cite{AMFlow,AMFlow2}.
The master integrals at other kinematic points are obtained by solving differential equations in the kinematic variables with \texttt{DESolver}, following the strategy discussed in Ref.~\cite{Liu:2023special}.

Then we calculate the counterterm contributions within the HEFT framework.
The masses and fields are renormalized in the on-shell scheme, and the electric charge is defined in the $G_\mu$ scheme, which requires computing NLO EW corrections to the $\mu$ decay width using $\mathcal{L}_{\rm HEFT}$.
We use the tadpole prescription with $\delta t=-T$, which completely cancels the tadpole-diagram contribution $T$, including the finite parts~\cite{Denner:1991kt}.
The modified couplings and the additional local couplings are renormalized using two different prescriptions. In the first prescription, we require the one-particle-irreducible (1PI) subdiagrams to be finite \cite{Herrero:2022krh}. Consequently, the ultraviolet divergence of each 1PI loop diagram is absorbed either into the renormalization of the parameters in the LO Lagrangian or into new operators in the NLO Lagrangian, which is a distinctive feature of renormalization in the HEFT.
At NLO, seven new operators are required, including \((\partial h)^2\Box h/v^3\) and \((-m_t\bar{Q}_L U e_t t_R+{\rm h.c.}) \Box h/v^3\). Since most of these new operators are present only under off-shell kinematic conditions, we refer to this strategy as the off-shell renormalization prescription.

In contrast, we have also implemented an alternative on-shell renormalization prescription, in which only the on-shell operators that cannot be eliminated by the equations of motion are retained. In practice, two additional operators are required, \((\partial h)^2(m_t\bar{Q}_L U e_t t_R+{\rm h.c.})/v^4\) and \(-\alpha_s h^2/(8\pi v^4)(D_{\rho}G^a_{\mu\nu})(D^{\rho}G^{a\mu\nu})\); these can be written as linear combinations of the operator basis of Refs.~\cite{Sun:2023NLO,Sun:2023NNLO} and have been studied in Ref.~\cite{Brivio:2026HEFT}.
In this case, the renormalization of each coupling modifier or additional local coupling is performed by demanding the finiteness of a process with appropriate external states at an on-shell kinematic point, which is not necessarily a physical point. 
This procedure is carried out from lower-point to higher-point interactions and from lower to higher loop orders, until the two-loop \(gg\to hh\) process is reached. 
We have compared the resulting counterterms with those obtained using the background-field method~\cite{Buchalla:2018Renormalization} where possible, and found agreement.

In both prescriptions, the \(\overline{\mathrm{MS}}\) scheme is applied to renormalize the coupling modifiers and additional local couplings. Consequently, these are scale-dependent parameters. Including the scale-running effects of these parameters cancels the EW scale dependence in the loop amplitudes through NLO.
In what follows, all results are expressed in terms of the parameters at the reference scale \(\mu_0=m_h\), i.e., \(\kappa_i\equiv\kappa_i(\mu_0)\) and \(c_i\equiv c_i(\mu_0)\). In this work, we assume that the couplings of the additional operators, in both the off-shell and on-shell prescriptions, that are introduced solely for renormalization are set to zero, since they are absent in the SM and enter the amplitude only at EW NLO. Their contributions to \(gg\to hh\) can be computed from LO diagrams alone and are therefore straightforward to include.

One important subtlety about the renormalization in the HEFT is that the results for the mass and wave-function renormalization constants of the Higgs field, \(\delta m_h\) and \(\delta Z_h\), are different in the two prescriptions due to the presence of \(\mathcal{O}(p^4)\) terms in the divergences of the Higgs self-energy diagrams. In the off-shell prescription, we add a new off-shell operator, \((\Box h)^2/v^2\), to cancel these divergences before applying the on-shell renormalization scheme. No such operator is introduced in the on-shell prescription. However, we find that the two prescriptions yield the same result for the sum of all counterterm contributions.

Combining the two-loop amplitudes and the counterterm contributions, and performing the convolution with parton distribution functions (PDFs) and phase space integrations, we obtain the cross section as a general function of the coupling modifiers ($\kappa_i$) and additional local couplings ($c_i$).
We perform the calculations in both the Feynman and unitary gauges, reaching the same result.
In fact, each coefficient of the coupling monomial in the cross section is gauge-independent. As a special case, our result with \(\kappa_i=1\) and \(c_i=0\) provides an independent cross-check of the complete SM NLO EW result of Ref.~\cite{Bi:2024}, which was obtained only in the Feynman gauge.


In the numerical calculation,
we take $m_h=125\,\mathrm{GeV}$ and the mass ratios
\begin{equation}
 \frac{m_t^2}{m_h^2}=\frac{23}{12},\qquad
 \frac{m_W^2}{m_h^2}=\frac{12}{29},\qquad
 \frac{m_Z^2}{m_h^2}=\frac{17}{32}.
\end{equation}
The electroweak input is $G_\mu=1.16637\times10^{-5}\,\mathrm{GeV}^{-2}$.
We use the central PDF4LHC21\_40 PDF set~\cite{PDF4LHC21} and its associated strong coupling through \texttt{LHAPDF6}~\cite{LHAPDF}.
The QCD renormalization and factorization scales are set to $\mu_R=\mu_F=m_{hh}/2$.

At the 13 TeV LHC, the inclusive LO cross section can be written as
\begin{equation}
\begin{aligned}[b]
&\sigma_{\LO}=\Bigl[
\kappa_3^2\bigl(4.637\kappa_t^2+9.360\kappa_t c_{ggh}
+5.246c_{ggh}^2\bigr)
\\&+\kappa_3\bigl(-22.56\kappa_t^3-19.33\kappa_t^2c_{ggh}+46.41\kappa_t c_{2t}\\&+41.24\kappa_t c_{gghh}
+41.24c_{2t}c_{ggh}+47.92c_{ggh}c_{gghh}\bigr)
\\
&+34.37\kappa_t^4-139.9\kappa_t^2c_{2t}-91.84\kappa_t^2c_{gghh}
+167.0c_{2t}^2\\&+242.5c_{2t}c_{gghh}
+197.6c_{gghh}^2\Bigr]\,\mathrm{fb}.
\end{aligned}
\label{eq:lopoly}
\end{equation}
This result agrees with the calculation of~\cite{Heinrich:2020} when the same settings are used.
The EW correction is given by
\begin{align}
    \frac{\Delta\sigma_{\rm EW}}{\rm 10^{-3}fb}=\sum_{ijk\in \mathcal{I}}c_{2t}^i c_{ggh}^j c_{gghh}^k P_{ijk}(\kappa_3,\kappa_4,\kappa_t,\kappa_{V},\kappa_{2V})
\end{align}
where the \(P_{ijk}\) are polynomials in the coupling modifiers and the set $\mathcal{I}=$ $\{$000,\,100,\,010,\,001,\,200,\,110,\,101,\,020,
\,011,\,002,\,300,\,210,\,201,\,120,\,111$\}$.
The explicit form of each \(P_{ijk}\) can be found in the ancillary file.
The first term reads
\begin{widetext}
\begin{equation}
\begin{aligned}[b]
&P_{000}=\kappa_3^3\kappa_t^2(65.97\kappa_3-79.56\kappa_t-22.69)
+\kappa_3^2\kappa_t\bigl[\kappa_t(344.1\kappa_t^2+288.2\kappa_t
-205.2+15.12\kappa_V^2-31.26\kappa_V\\
&{}+16.13\kappa_{2V}-5.556\kappa_4)-118.9\kappa_V\bigr]
+\kappa_3\kappa_t\bigl[\kappa_t^2(-2467\kappa_t^2-566.8\kappa_t
+41.43+119.6\kappa_V^2+76.05\kappa_V\\
&{}-2.456\kappa_{2V}+68.90\kappa_4)
+\kappa_t\kappa_V(811.2+363.7\kappa_{2V}-426.0\kappa_V^2)
-659.8\kappa_{2V}+1244\kappa_V^2\bigr]
+\kappa_t^2\bigl[\kappa_t^2(2908\kappa_t^2\\
&{}+887.7+168.3\kappa_V^2-112.2\kappa_{2V}-145.9\kappa_4)
+\kappa_t\kappa_V(-1082+326.5\kappa_{2V}-790.0\kappa_V^2)
+1188\kappa_{2V}-2894\kappa_V^2\bigr].
\label{eq:p000}
\end{aligned}
\end{equation}
\end{widetext}

\begin{figure*}[t]
\centering
\includegraphics[width=0.8\textwidth]{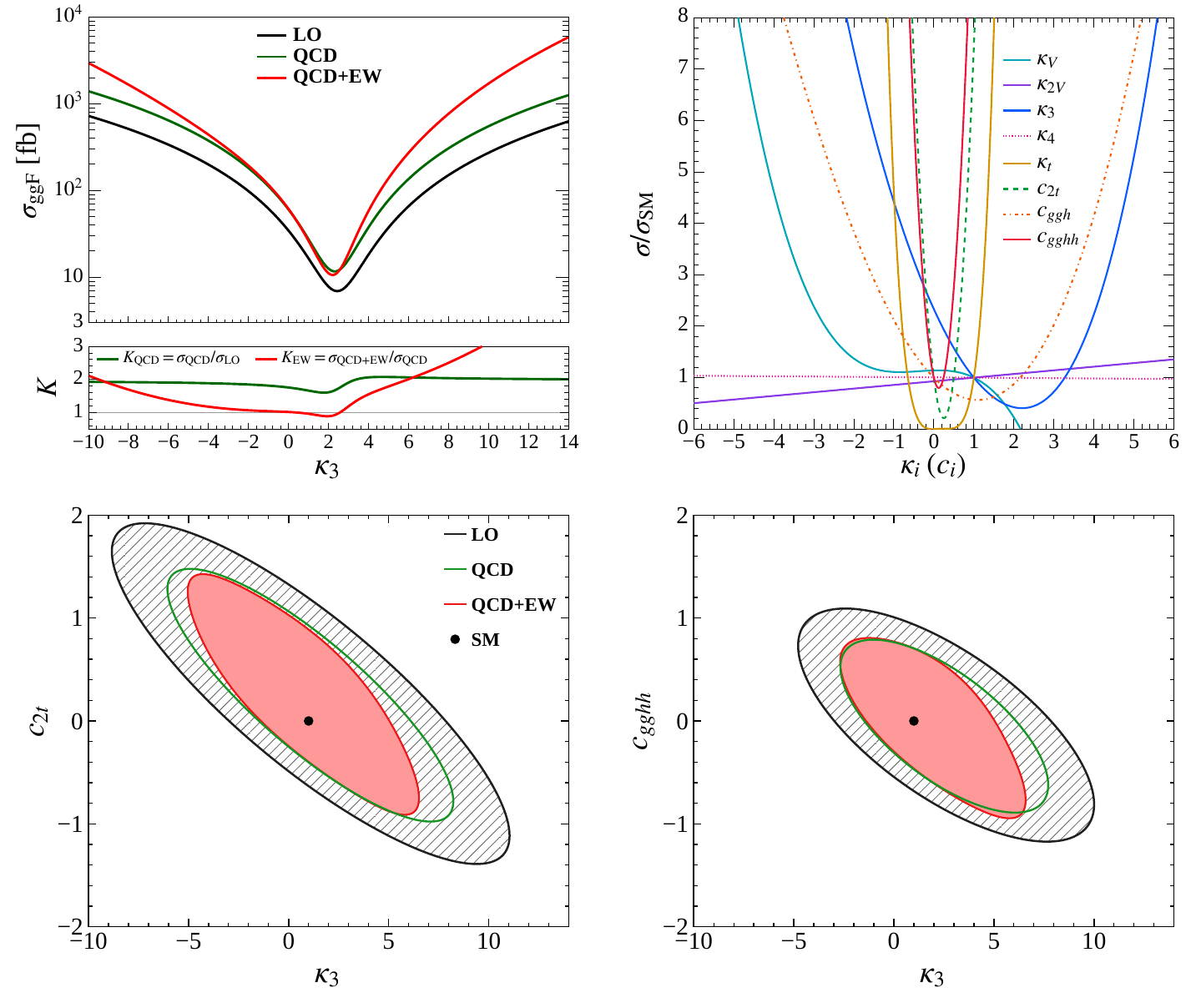}
\caption{
Cross sections of Higgs-pair production via gluon fusion at the 13 TeV LHC.
Upper-left: The cross sections as functions of \(\kappa_3\) at different perturbative orders.
Upper-right: The normalized cross section with NLO QCD and EW corrections as a function of a single parameter.
When one parameter is varied, the others are fixed at their SM values.
Lower-left: Allowed regions in the \(\kappa_3\)--\(c_{2t}\) plane.
Lower-right: Allowed regions in the \(\kappa_3\)--\(c_{gghh}\) plane.
 }
\label{fig:rates}
\end{figure*}
\begin{figure*}[t]
\centering
\includegraphics[width=0.9\textwidth]{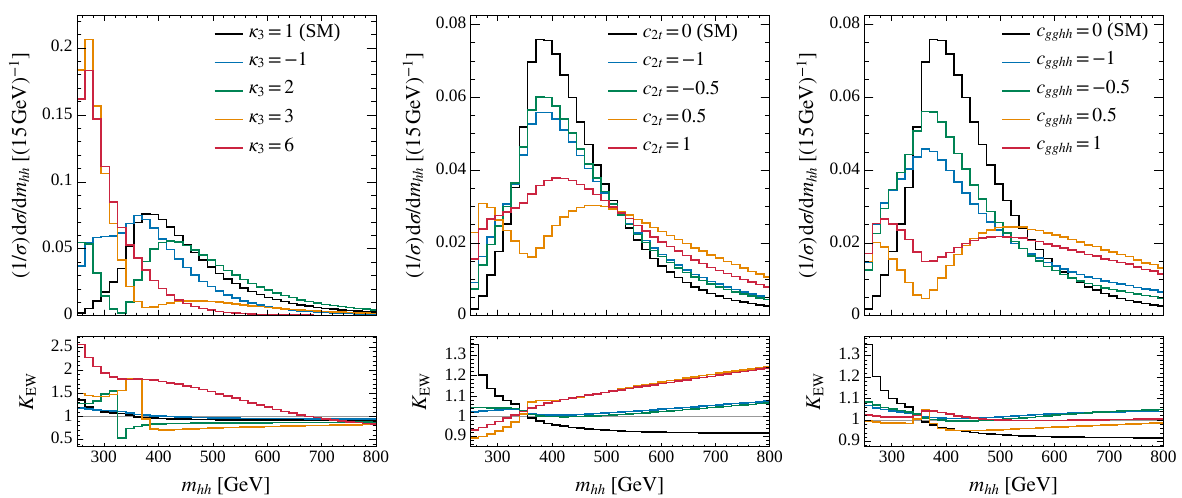}
\caption{Normalized Higgs-pair invariant mass distributions at the 13 TeV LHC with both NLO QCD and EW corrections for different values of $\kappa_3$ (left), $c_{2t}$ (middle), and $c_{gghh}$ (right).
The lower panels show the $K_{\rm EW}$ factor.}
\label{fig:spectra}
\end{figure*}

We find that the dependence on the Higgs trilinear self-coupling modifier \(\kappa_3\) and the top-Yukawa coupling modifier \(\kappa_t\) extends up to the fourth and sixth powers, respectively, indicating a strong sensitivity to these parameters. The \(\kappa_V^3\) term arises from the correction to the \(hhh\) vertex through a vector-boson loop. The quartic Higgs self-coupling modifier \(\kappa_4\) and the \(VVhh\) coupling modifier \(\kappa_{2V}\) enter only linearly at this order.
The top-pair--Higgs-pair local coupling \(c_{2t}\) enters the cross section cubically, which illustrates the advantage of using \(gg\to hh\) to probe this coupling. The gluon--Higgs local couplings \(c_{ggh}\) and \(c_{gghh}\) enter quadratically. The equations above reveal intricate correlations among the coupling modifiers and local couplings.

To better understand the dependence on each parameter, we show in Fig.~\ref{fig:rates} the cross sections obtained when only one parameter is varied while the others are fixed to their SM values. The upper-left panel displays the cross sections at different perturbative orders as \(\kappa_3\) is varied. The LO curve is obtained from Eq.~(\ref{eq:lopoly}), while the NLO QCD result is computed using the \texttt{ggHH} implementation~\cite{Heinrich:2017kxx,Heinrich:2019bkc,Heinrich:2020} in \texttt{POWHEG-BOX-V2}~\cite{Nason:2004rx,Frixione:2007vw,Alioli:2010xd}.
We combine the EW corrections with a QCD baseline using
\begin{equation}
 \sigma_{\QCD+\EW}=
 \sigma_{\QCD}
 \left(1+\frac{\Delta\sigma_{\EW}}
 {\sigma_{\LO}}\right).
 \label{eq:combine}
\end{equation}
This prescription provides an approximate treatment of mixed QCD--EW effects by combining the separate QCD and EW corrections multiplicatively. This factorized form has been shown to provide a good approximation for \(gg\to h\)~\cite{Bonetti:2018ggH}.
This plot shows that the QCD corrections hardly change the parabolic shape. In contrast, the EW corrections, which introduce additional higher-power terms, significantly increase the cross section for \(|\kappa_3|>3\) and also slightly shift the position of the minimum.
The \(K_{\rm EW}\) factor, defined as \(1+\Delta\sigma_{\rm EW}/\sigma_{\rm LO}\), is 0.96 at \(\kappa_3=1\), consistent with the SM result in Ref.~\cite{Bi:2024}. It increases to 1.19 at \(\kappa_3=3\) and reaches 2.03 at the current experimental upper bound \(\kappa_3=6.1\), where the EW correction is comparable to the QCD correction. This bound was obtained using predictions with higher-order QCD corrections but without EW corrections~\cite{ATLASCMS:2026}. Therefore, including the EW corrections for general values of \(\kappa_3\) should significantly tighten the constraint on \(\kappa_3\).

The upper-right panel of Fig.~\ref{fig:rates} shows the cross section including both NLO QCD and EW corrections, normalized to the SM result, as a function of each coupling modifier or additional local coupling. The cross section is most sensitive to the local couplings \(c_{2t}\) and \(c_{gghh}\), and to the top-Yukawa coupling modifier \(\kappa_t\).
Note that \(c_{2t}\) and \(c_{gghh}\) can enter the single-Higgs production cross section via EW corrections, while \(\kappa_t\) appears at LO. However, the sensitivity may be weaker since only lower-power terms are involved there. We also observe that the cross section changes slowly with respect to \(\kappa_4\) and \(\kappa_{2V}\). Triple-Higgs and vector-boson-fusion double-Higgs production are likely better channels to constrain these two parameters.

There are strong correlations among the parameters. In the lower panel of Fig.~\ref{fig:rates}, we show two examples of such correlations. We take the experimental upper limit of the gluon-fusion Higgs-pair cross section to be 142.7 fb, so that the same upper bound on \(\kappa_3\) is reproduced using the QCD-only prediction.
As $c_{2t}$ increases, the allowed region of $\kappa_3$ shifts toward lower values.
The EW corrections markedly change the shape of the allowed region in the $\kappa_3$--$c_{2t}$ plane, particularly at large $|\kappa_3|$.
The EW corrections have a similar effect on the correlation between $\kappa_3$ and $c_{gghh}$.

In addition to the total cross section, the features of the kinematic distributions can be used to design dedicated event selection cuts that enhance the signal and suppress the backgrounds.
Figure~\ref{fig:spectra} shows the normalized distributions of the Higgs-pair invariant mass $m_{hh}$ at the 13 TeV LHC, including the NLO QCD and EW corrections, for variations in $\kappa_3, c_{2t}$ and $c_{gghh}$, respectively.
For $\kappa_3=1$, the $m_{hh}$ distribution peaks around 375 GeV.
The $K_{\rm EW}$ factor decreases from 1.35 in the first bin to 0.92 at large $m_{hh}$.
This behaviour is consistent with the SM result reported in Ref. \cite{Bi:2024}.
When $\kappa_3$ is changed to $-1$, the peak shifts toward lower $m_{hh}$ and the distribution at low $m_{hh}$ is enhanced.
The corresponding $K_{\rm EW}$ factor is 1.18 at threshold and decreases with increasing $m_{hh}$.
In the other direction, when $\kappa_3$ is increased to 2, 
the shape of the distribution changes dramatically.
A dip appears around $m_{hh}=325$ GeV.
The $K_{\rm EW}$ factor changes abruptly at this point, dropping from 1.5 to 0.5.
For larger \(\kappa_3\), the cross section receives its dominant contributions at small \(m_{hh}\). The EW corrections are more pronounced around the peak, unlike in the SM.

The invariant mass distribution is also sensitive to \(c_{2t}\), as shown in the middle panel of Fig.~\ref{fig:spectra}. For \(c_{2t}=-0.5\), the distribution is broader than in the SM case.
The EW corrections are small in the peak region and increase to 7\% at large $m_{hh}$.
For \(c_{2t}=0.5\), two peaks appear, one near the Higgs-pair threshold and the other at 480 GeV, and the spectrum at large \(m_{hh}\) falls off more slowly than in the SM. The \(K_{\rm EW}\) factor increases monotonically from 0.89 to 1.24.

Varying \(c_{gghh}\) significantly affects the \(m_{hh}\) distribution, as is evident from the right panel of Fig.~\ref{fig:spectra}. For negative \(c_{gghh}\), the peak position shifts toward lower \(m_{hh}\). The EW corrections first decrease and then increase as \(m_{hh}\) increases, but remain below 9\%.
For positive values of \(c_{gghh}\), a bimodal structure appears, and the EW corrections are more pronounced in the trough region. In all cases, the EW corrections cannot be captured by a constant rescaling of the distribution.

In summary, we have presented the full EW corrections to gluon-fusion Higgs-pair production within the HEFT, in which the Higgs self-couplings and the Higgs couplings with other particles can vary freely. 
We have calculated the massive two-loop amplitudes using a range of modern multiloop techniques and have devised two independent prescriptions to address the EW renormalization of the coupling modifiers and additional local couplings.
Calculations in the Feynman and unitary gauges provide complementary checks, and the SM limit provides a cross-check of the existing complete NLO EW result. Numerical results show that the EW corrections significantly enhance the cross section for \(|\kappa_3|>3\) and visibly change the correlation between \(c_{2t}\) (\(c_{gghh}\)) and \(\kappa_3\).
The shape of the Higgs-pair invariant mass distribution is sensitive to the values of \(\kappa_3\), \(c_{2t}\), and \(c_{gghh}\), and the EW corrections are sizable in many parameter scenarios and kinematic regions. These results demonstrate the need for coupling-dependent EW predictions in precision studies of Higgs-pair production at the LHC.

To compare with the experimental data at the LHC, we need to combine the theoretical predictions for the gluon-fusion and vector-boson-fusion Higgs-pair production channels.
Including Higgs decays~\cite{Li:2024bbgg,Li:2025bbtautau} and parton showers~\cite{Braun:2025HHshower} is also a necessary step toward fully simulating the signal events.
The methodology developed in this work, in particular the EW renormalization within the HEFT, can be readily extended to these processes.
Taken together, these ingredients would provide a precise theoretical basis for understanding the core of the Higgs mechanism---the Higgs potential.

We would like to thank Huan-Yu Bi and Yan-Qing Ma for providing the SM result for comparison.
We also thank Hong-Yang Han and Hai Tao Li for helpful discussions.
This work was supported by the National Natural Science Foundation of China under Grants No. 12375076  and No. 12321005.

\end{document}